\documentclass[conference]{IEEEtran}
\IEEEoverridecommandlockouts
\title{Minimax Data Sanitization with Distortion Constraint and Adversarial Inference}

\author{\IEEEauthorblockN{Amirarsalan Moatazedian\IEEEauthorrefmark{1}, Yauhen Yakimenka\IEEEauthorrefmark{1}, R\'{e}mi A. Chou\IEEEauthorrefmark{2}, and J{\"o}rg Kliewer\IEEEauthorrefmark{1}}
\IEEEauthorblockA{\IEEEauthorrefmark{1}Helen and John C.~Hartmann Department of Electrical and Computer Engineering,\\ New Jersey Institute of Technology, Newark, New Jersey 07102, USA
}\IEEEauthorblockA{\IEEEauthorrefmark{2}University of Texas at Arlington, Arlington, TX 76019, USA \\ 
Email: \{am3734, yauhen.yakimenka, jkliewer\}@njit.edu, remi.chou@uta.edu
}
\thanks{This work was supported by U.S. National Science Foundation grants CCF-2201824 and CCF-2425371.}
}

\usepackage{amsmath, amsfonts, amssymb}
\usepackage{amsthm}          
\usepackage{graphicx}
\usepackage{mathtools}
\usepackage{breqn}
\usepackage{listings}
\usepackage{float}
\usepackage[dvipsnames]{xcolor}
\usepackage[normalem]{ulem}    
\usepackage{comment}
\usepackage{newunicodechar}
\newunicodechar{ }{\,}
\usepackage{hyperref}
\usepackage[capitalize]{cleveref}
\crefname{equation}{}{}
\usepackage{tikz}
\usepackage{pgfplots}
\usepackage{booktabs}

\usetikzlibrary{arrows.meta, positioning}

\pgfplotsset{
    every axis/.style={
        width=0.8\linewidth,          
        height=0.6\linewidth,         
        xlabel={\fontsize{8pt}{9.6pt}\selectfont X-Axis},
        ylabel={\fontsize{8pt}{9.6pt}\selectfont Y-Axis},
        label style={font=\fontsize{8pt}{9.6pt}\selectfont},
        tick label style={font=\fontsize{8pt}{9.6pt}\selectfont},
        legend style={
            font=\fontsize{8pt}{9.6pt}\selectfont,
            at={(0.02,0.98)},
            anchor=north west
        },
        title style={
            font=\fontsize{8pt}{9.6pt}\selectfont,
            align=center, yshift=-5pt
        },
        legend cell align={left},
        scale only axis,
    },
    compat=1.17
}

\usetikzlibrary{plotmarks}

\usepackage[linesnumbered,ruled,vlined]{algorithm2e}

\SetAlgoNlRelativeSize{-1} 
\SetAlFnt{\small} 
\SetKwInput{KwIn}{Input}
\SetKwInput{KwOut}{Output}

\definecolor{arsa}{rgb}{1.0, 0.8, 0.6} 
\definecolor{myblue}{RGB}{0, 102, 204}

\newtheorem{proposition}{Proposition}

\lstdefinestyle{mystyle}{
    basicstyle=\ttfamily\footnotesize,
    breaklines=true,
    numbers=left,
    numberstyle=\tiny,
    stepnumber=1,
    numbersep=5pt,
    frame=single,
    captionpos=b,
    keywordstyle=\color{blue},
    commentstyle=\color{green!50!black},
    stringstyle=\color{orange},
    showspaces=false,
    showstringspaces=false,
    showtabs=false,
    tabsize=2,
    language=Python
}
\renewcommand{\Pr}{\mathbb{P}}

\newcommand{\EE}{\mathbb{E}}
\DeclareMathOperator{\Var}{Var}

\begin{document}

\maketitle

\begin{abstract}
We study a privacy-preserving data-sharing setting where a privatizer transforms private data into a sanitized version observed by an authorized reconstructor and two unauthorized adversaries, each with access to side information correlated with the private data. 

The reconstructor is evaluated under a distortion function, while each adversary is evaluated using a separate loss function. The privatizer ensures the reconstructor distortion remains below a fixed threshold while maximizing the minimum loss across the two adversaries. This two-adversary setting models cases where individual users cannot reconstruct the data accurately, but their combined side information enables estimation within the distortion threshold. The privatizer maximizes individual loss while permitting accurate reconstruction only through collaboration. This echoes secret-sharing principles, but with lossy rather than perfect recovery. We frame this as a constrained data-driven minimax optimization problem and propose a data-driven training procedure that alternately updates the privatizer, reconstructor, and adversaries. We also analyze the Gaussian and binary cases as special scenarios where optimal solutions can be obtained. These theoretical optimal results are benchmarks for evaluating the proposed minimax training approach.
\end{abstract}


\section{Introduction}

In distributed data-sharing, users often hold distinct pieces of side information correlated with a private dataset~\cite{villard2010secure, ekrem2011secure, sankar2013utility}. Prior work has studied how to release sanitized data that enables useful reconstruction at a legitimate user while limiting information leakage to unauthorized users. For example, secure source coding frameworks~\cite{villard2010secure, ekrem2011secure} examine tradeoffs among rate, distortion, and leakage, while database sanitization approaches~\cite{sankar2013utility} model privacy using entropy-based metrics and utility via distortion. More recent formulations consider access structures to determine which coalition of users can successfully decode under distortion constraints~\cite{zivarifard2024secure}.

We study a related but distinct setting where a privatizer releases a sanitized version of private data to one reconstructor and two adversaries, each with separate side information. The reconstructor and adversaries attempt to estimate the private data using the sanitized output and their side information. We formalize this as a constrained minimax problem: the reconstructor minimizes its distortion, which determines utility; each adversary minimizes its loss, which determines privacy; and the privatizer maximizes the minimum adversarial loss subject to a constraint on reconstructor distortion. This induces a utility–privacy tradeoff shaped by the privatizer’s strategy and the distribution of side information. For Gaussian and binary data, we derive closed-form or piecewise-linear program solutions showing how relaxed utility constraints improve privacy. We also propose a data-driven minimax procedure not restricted to any specific distribution, distortion, or loss functions. This procedure alternates updates of the privatizer, reconstructor, and adversaries. 

While our objective and experiments treat the adversaries separately, always maximizing the minimum adversarial loss across the two, the system structure permits scenarios where utility is granted only to a coalition. That is, although no individual user is trusted to reconstruct within the distortion threshold, their combined side information may suffice. The presence of the reconstructor as a separate entity in our model allows an abstract modeling of a coalition of unauthorized users. This setup connects to threshold reconstruction and secret-sharing principles, but under approximate rather than exact recovery.


Threshold-based reconstruction is a fundamental concept in cryptographic protocols, where a prescribed number of participants must combine their respective shares for successful recovery. Classical secret sharing schemes~\cite{shamir1979share,Blakley1899SafeguardingCK} assume a secure share distribution mechanism and guarantee exact reconstruction for qualified subsets, while ensuring that unauthorized subsets gain no information about the secret. Our model instead makes sanitized data publicly available to both the reconstructor and the adversaries, who attempt to minimize their respective distortion and losses using the sanitized output along with their side information. As a result, exact recovery and perfect privacy are no longer possible, and each party incurs an estimation loss.

Several works on secure source coding, such as~\cite{villard2010secure, ekrem2011secure}, measure utility via distortion at a legitimate decoder and evaluate privacy through equivocation or mutual information at an adversary, and extensions like~\cite{zivarifard2024secure} include access structures for reconstruction. In database sanitization~\cite{sankar2013utility}, utility is modeled as distortion in public attributes while privacy is given by the conditional entropy of private attributes. We similarly balance utility and privacy but rely on both a reconstructor distortion and adversarial loss. Our formulation frames privacy as estimation loss, rather than information-theoretic uncertainty, and uses an iterative minimax procedure where the privatizer produces sanitized data subject to a constraint on an authorized reconstructor distortion while maximizing the minimum adversarial loss across two unauthorized adversaries.

Adversarial learning frameworks like context-aware generative adversarial privacy (GAP)~\cite{huang2017context} also frame privacy as a minimax objective and include both analytical and data-driven methods. However, our formulation differs from GAP in that, unlike GAP, our model explicitly includes a separate reconstructor that aims to minimize distortion in estimating the private variable, using side information and the sanitized output. This separation reflects practical scenarios where sanitized data is released to external users whose behavior cannot be coordinated with the privatizer.


The rest of the paper is organized as follows. Section~\ref{sec:system_model} defines the system model and optimization objective. Section~\ref{sec:analytical_solutions} presents analytical results for two data models. Section~\ref{sec:data_driven_approach} introduces our neural-network-based learning method. Section~\ref{sec:simulation_results} reports empirical results, and Section~\ref{sec:conclusion} concludes the paper.


\subsection*{Notation}
Throughout this paper, random variables are denoted by uppercase letters (e.g., \(X\)), their realizations by lowercase letters (e.g., \(x\)), and their domains by calligraphic letters (e.g., \(\mathcal{X}\)). Probabilities and expectations are denoted by \(\Pr(\cdot)\) and \(\EE[\cdot]\), respectively. We use \(\Var[\cdot]\) to denote variance. The positive-part operator is defined as \((u)^+ = \max\{u,0\}\).


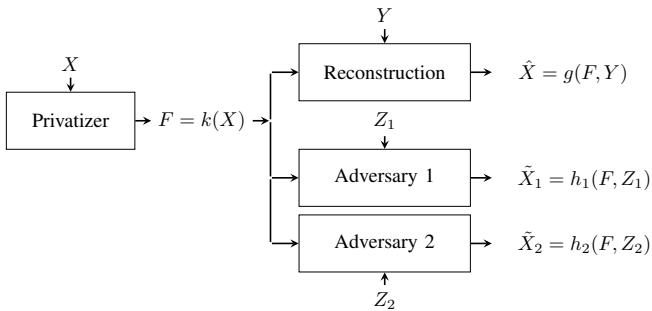
\begin{figure}[t]\centering
\resizebox{\columnwidth}{!}{
\begin{tikzpicture}[auto, node distance=2.5cm, >=stealth]
\node [align=center, inner sep=0pt] (center) {};
\node [above of=center, node distance=0.85cm, align=center, inner sep=0pt] (rleft) {};
\node [below of=center, node distance=1cm, align=center, inner sep=0pt] (adv1left) {};
\node [below of=center, node distance=2.15cm, align=center, inner sep=0pt] (adv2left) {};
\node [left of=center, node distance=3.5cm, draw, rectangle, minimum width=2.25cm, minimum height=1cm, align=center] (privatizer) {Privatizer};
\node [above of=privatizer, node distance=1cm] (x) {\(X\)};
\node [right of=rleft, node distance=2cm, draw, rectangle, minimum width=3cm, minimum height=1cm, align=center] (reconstruction) {Reconstruction};
\node [right of=adv1left, node distance=2cm, draw, rectangle, minimum width=3cm, minimum height=1cm, align=center] (adversary1) {Adversary 1};
\node [right of=adv2left, node distance=2cm, draw, rectangle, minimum width=3cm, minimum height=1cm, align=center] (adversary2) {Adversary 2};
\node [right of=reconstruction, node distance=2cm, align=center] (rright) {};
\node [right of=adversary1, node distance=2cm, align=center] (adv1right) {};
\node [right of=adversary2, node distance=2cm, align=center] (adv2right) {};
\node [above of=reconstruction, node distance=1cm] (y) {\(Y\)};
\node [above of=adversary1, node distance=1cm] (z1) {\(Z_1\)};
\node [below of=adversary2, node distance=1cm] (z2) {\(Z_2\)};
\node [right of=rright, node distance=1.3cm] (xhat) {\(\hat{X}=g(F,Y)\)};
\node [right of=adv1right, node distance=1.5cm] (x1tilde) {\(\tilde{X}_1=h_1(F,Z_1)\)};
\node [right of=adv2right, node distance=1.5cm] (x2tilde) {\(\tilde{X}_2=h_2(F,Z_2)\)};
\node [left of=center, node distance=1.2cm] (fx) {\(F=k(X)\)};
\draw[->, thick, >=stealth] (x) -- (privatizer);
\draw[-, thick] (center) -- (rleft);
\draw[-, thick] (center) -- (adv1left);
\draw[-, thick] (adv1left) -- (adv2left);
\draw[->, thick, >=stealth] (rleft) -- (reconstruction);
\draw[->, thick, >=stealth] (adv1left) -- (adversary1);
\draw[->, thick, >=stealth] (adv2left) -- (adversary2);
\draw[->, thick, >=stealth] (y) -- (reconstruction);
\draw[->, thick, >=stealth] (z1) -- (adversary1);
\draw[->, thick, >=stealth] (z2) -- (adversary2);
\draw[->, thick, >=stealth] (reconstruction) -- (rright);
\draw[->, thick, >=stealth] (adversary1) -- (adv1right);
\draw[->, thick, >=stealth] (adversary2) -- (adv2right);
\draw[->, thick, >=stealth] (fx) -- (center);
\draw[->, thick, >=stealth] (privatizer) -- (fx);
\end{tikzpicture}
}
\caption{System model: The privatizer maps the private data \(X\) to a sanitized version \(F = k(X)\). The reconstructor estimates \(\hat{X} = g(F, Y)\), while adversaries estimate \(\tilde{X}_i = h_i(F, Z_i)\) for \(i = 1, 2\).}
\label{fig:privacy_framework}
\end{figure}


\section{System Model and Problem Formulation}\label{sec:system_model}

The overall system is illustrated in Figure~\ref{fig:privacy_framework}. Consider a private dataset consisting of i.i.d. random variables, with each entry modeled as \(X \in \mathcal{X}\). The privatizer observes the private data \(X\) and applies a mapping \(k : \mathcal{X} \to \mathcal{F}\) to produce a sanitized version \(F = k(X)\), which is made publicly available.
An authorized reconstructor, equipped with side information \(Y \in \mathcal{Y}\), attempts to estimate \(X\) using \(g : \mathcal{F} \times \mathcal{Y} \to \widehat{\mathcal{X}}\), producing \(\hat{X} = g(F, Y)\).

Two non-colluding adversaries, each with separate side information \(Z_1, Z_2 \in \mathcal{Z}\), attempt to infer \(X\) from \(F\) using \(h_i : \mathcal{F} \times \mathcal{Z} \to \widehat{\mathcal{X}}\), resulting in estimates \(\tilde{X}_i = h_i(F, Z_i)\) for \(i = 1, 2\). We assume the joint distribution \(P_{X,Y,Z_1,Z_2}\) is known. The privatizer only observes \(X\), not the side information. The mappings $k$, $g$, and $h_i$ are potentially randomized.

We use a distortion function \(d(X, \hat{X}): \mathcal{X} \times \mathcal{X} \to \mathbb{R}\) for the reconstructor, where smaller values indicate more accurate reconstruction. For each adversary \(i\), we define a loss function \(\ell(X, \tilde{X}_i): \mathcal{X} \times \mathcal{X} \to \mathbb{R}\). Since \(X\) is random, we measure distortion and loss using their expected values: \(\mathbb{E}[\,d(X, \hat{X})\,]\) for the reconstructor and \(\mathbb{E}[\,\ell(X, \tilde{X}_i)\,]\) for adversary~\(i\).

The objective is to maximize the minimum estimation loss across the two adversaries, while ensuring that the reconstructor expected distortion satisfies \(\mathbb{E}[\,d(X, \hat{X})\,] \le D\). Formally, we aim to solve the following minimax optimization problem:
\begin{equation}
\label{eq:optimization_problem}
\begin{aligned}
&\max_{k,g} 
\;\;\min_{\,h_1,\,h_2}\;\;\;
  \Bigl\{
     \mathbb{E}[\ell(X,\tilde{X}_1)],
     \,\mathbb{E}[\ell(X,\tilde{X}_2)]
  \Bigr\} \\[5pt]
&\quad\text{subject to}\quad 
 \mathbb{E}[\,d(X,\hat{X})\,]\;\le\;D,
\end{aligned}
\end{equation}
where the expectations are taken over \(P_{X,Y,Z_1,Z_2}\) and any internal randomness in \(k\), \(g\), \(h_1\), and \(h_2\). Note that \eqref{eq:optimization_problem} entails two contradicting perspectives: on the one hand, we aim to keep reconstructor distortion below a threshold $D$, on the other hand, the minimum adversarial loss is maximized.


We now describe a special case of the general setup, where the reconstructor is not a separate entity, but rather an abstract representation of a coalition formed by users with partial side information. The privatizer is designed to ensure that the combined side information \(Y = (Z_1, Z_2)\), together with \(F\), enables reconstruction within the distortion threshold, while also maximizing the expected loss when only one side variable \(Z_i\) is used to estimate \(\tilde{X}_i = h_i(F, Z_i)\), consistent with the requirement in~\eqref{eq:optimization_problem}. This setting relates to the principle behind secret-sharing: only when enough partial information is pooled (here modeled by \((Z_1, Z_2)\)) can the desired output be accurately inferred. However, unlike classical schemes that guarantee perfect recovery~\cite{shamir1979share,Blakley1899SafeguardingCK}, we allow for approximate reconstruction with distortion no greater than \(D\), and focus on maximizing the loss incurred by any user relying on limited side information.


\section{Gaussian and Binary Optimal Solutions}
\label{sec:analytical_solutions}

We now present two concrete data models, one Gaussian and one binary, for which we can derive optimal solutions of \eqref{eq:optimization_problem}. These solutions also serve as benchmarks for our data-driven minimax approach (see \cref{sec:data_driven_approach}).


\subsection{Gaussian Data Model}
\label{subsec:gaussian_data}

Assume the private variable \(X\) and the side information \((Y, Z_1, Z_2)\) follow a joint Gaussian distribution with means \(\mu_X, \mu_Y, \mu_{Z_1}, \mu_{Z_2}\), variances \(\sigma_X^2, \sigma_Y^2, \sigma_{Z_1}^2, \sigma_{Z_2}^2\), and correlations \(\rho_{XY}, \rho_{XZ_1}, \rho_{XZ_2}\), etc. We use squared error~\cite[Ch.~11.4]{kay1993statistical} for both the reconstructor distortion and the adversarial loss:
\[
d(x, y) = \ell(x, y) = (x - y)^2,
\]
with mean squared error (MSE) defined as the expected value. We represent the joint statistics using the covariance matrix:
\BlankLine
\BlankLine
\scalebox{0.925}[1.0]{
$\displaystyle
\begin{bmatrix}
\sigma_X^2 & \rho_{XY}\sigma_X\sigma_Y & \rho_{XZ_1}\sigma_X\sigma_{Z_1} & \rho_{XZ_2}\sigma_X\sigma_{Z_2} \\
\rho_{XY}\sigma_X\sigma_Y & \sigma_Y^2 & \rho_{YZ_1}\sigma_Y\sigma_{Z_1} & \rho_{YZ_2}\sigma_Y\sigma_{Z_2} \\
\rho_{XZ_1}\sigma_X\sigma_{Z_1} & \rho_{YZ_1}\sigma_Y\sigma_{Z_1} & \sigma_{Z_1}^2 & \rho_{Z_1Z_2}\sigma_{Z_1}\sigma_{Z_2} \\
\rho_{XZ_2}\sigma_X\sigma_{Z_2} & \rho_{YZ_2}\sigma_Y\sigma_{Z_2} & \rho_{Z_1Z_2}\sigma_{Z_1}\sigma_{Z_2} & \sigma_{Z_2}^2
\end{bmatrix}.
$}
\BlankLine
\BlankLine

We assume the privatizer outputs a variable \(F\) jointly Gaussian with \(X\), e.g., by adding independent Gaussian noise. This design allows for selecting the correlation between \(X\) and \(F\). The optimal MSE estimators~\cite[Ch.~11.4]{kay1993statistical} are:
\[
\hat{X} = \EE[X | F, Y] \quad \text{and} \quad \tilde{X}_i = \EE[X | F, Z_i].
\]
The reconstructor distortion and the adversarial loss~\cite[Ch.~11.4]{kay1993statistical} are respectively given by
\[
\EE[\Var[X | F, Y]] \quad \text{and} \quad \EE[\Var[X | F, Z_i]].
\]
By the law of total variance~\cite[Ch.~4]{casella2024statistical}, 
\[
\Var[X | F, Y] \le \Var[X | Y] \;\; \text{and} \;\; \Var[X | F, Z_i] \le \Var[X | Z_i].
\] 
These upper bounds correspond to the optimal estimation performance achievable using only the respective side information, which remains accessible regardless of the privatizer.
\begin{proposition}[Proof in Appendix~\ref{app:gaussian}]
\label{prop:gaussian_conditional_var}
In the jointly Gaussian case, with MSE as the metric for the reconstructor distortion and adversarial loss, the optimal solution of~\eqref{eq:optimization_problem} is 
\[
\min
\Bigl\{\frac{D'}{1 + Q_1 D'},\frac{D'}{1 + Q_2 D'}
\Bigr\},
\]
where 
\begin{align*}
Q_i &=
\frac{\rho_{XZ_i}^2 - \rho_{XY}^2}{\sigma_X^2\,
\bigl(1 - \rho_{XZ_i}^2\bigr)\bigl(1 - \rho_{XY}^2\bigr)}, \\
D'&=\min
\{D, \, \EE[\Var[X | Y]]\}.
\end{align*}
\end{proposition}


\subsection{Binary Data Model}
\label{subsec:binary_data}

Consider a Bernoulli private variable \(X\in\{0,1\}\) with the probability of success \(\Pr(X=1)=p\). The reconstructor’s side information \(Y\) and each adversary’s side information \(Z_i\) are outputs of independent binary symmetric channels with crossover probabilities \(q_Y\) and \(q_{Z_i}\), respectively~\cite[Ch.~7]{cover2006elements}. We measure distortion and loss by Hamming distance (0-1 loss):
\[
d(x,y) = \ell(x,y) = \mathbf 1(x \neq y).
\]
The expected value of Hamming distance is misclassification probability, and it is minimized by the maximum a posteriori (MAP) estimator~\cite[Ch.~11.5]{kay1993statistical}. Let the privatizer flip \(X\) to produce \(F\in\{0,1\}\) with 
\[
s_0 = \Pr(F=0 | X=0),\quad
s_1 = \Pr(F=1 | X=1).
\]
Here, $s_0$ and $s_1$ are the design choices of the privatizer. One can use Proposition \ref{prop:binary_lp_solution} to find the optimal solution of~\eqref{eq:optimization_problem}.

\begin{proposition}[Proof in Appendix~\ref{app:binary}]
\label{prop:binary_lp_solution}
In the binary setting with 0-1 loss, the solution to the minimax problem~\eqref{eq:optimization_problem} is given by the piecewise-linear program:
\[
\begin{aligned}
\max_{\gamma,\, s_0,\, s_1} \quad & \gamma \\
\text{subject to} \quad 
& \mathcal{L}_{\mathrm{adv}_i}(s_0, s_1) \ge \gamma, \quad i = 1, 2, \\
& \mathcal{L}_{\mathrm{recon}}(s_0, s_1) \le D, \\
& 0 \le s_0, s_1 \le 1,
\end{aligned}
\]
where
\[
\begin{aligned}
\mathcal{L}_{\mathrm{recon}}(s_0, s_1) =\; & 
\min\left\{ (1 - p)s_0(1 - q_Y),\; p(1 - s_1)q_Y \right\} + \\
& \min\left\{ (1 - p)s_0 q_Y,\; p(1 - s_1)(1 - q_Y) \right\} + \\
& \min\left\{ (1 - p)(1 - s_0)(1 - q_Y),\; p s_1 q_Y \right\} + \\
& \min\left\{ (1 - p)(1 - s_0) q_Y,\; p s_1 (1 - q_Y) \right\},
\end{aligned}
\]
and the adversarial loss \(\mathcal{L}_{\mathrm{adv}_i}(s_0, s_1)\) is defined identically, except with \(q_Y\) replaced by \(q_{Z_i}\) in each term.
\end{proposition}

The piecewise-linear program can be reformulated as a mixed-integer program (MIP) and solved using MIP solvers.


\section{Data-Driven Approach}
\label{sec:data_driven_approach}

{
\setlength{\textfloatsep}{2pt plus 1pt minus 1pt}
\begin{algorithm}[t]
\caption{Data-Driven Minimax Training}
\label{alg:1}
\SetAlgoLined
\LinesNumbered
\DontPrintSemicolon
\SetKwInOut{Input}{Input}
\SetKwInOut{Output}{Output}

\Input{Dataset $\mathcal{D} = \{(x_j, y_j, z_{1,j}, z_{2,j}, r_j)\}_{j=1}^N$, distortion threshold $D$, number of epochs $T$, batch size $M$, initial parameters $\theta_k,\theta_g,\theta_{h_1},\theta_{h_2}$, penalty weight $\rho$, learning rates $\alpha_k,\alpha_g,\alpha_{h_1},\alpha_{h_2}$.} \nllabel{line:input}

\Output{Trained models $\theta_k,\theta_g,\theta_{h_1},\theta_{h_2}$.} \nllabel{line:output}

Initialize $\theta_k,\theta_g,\theta_{h_1},\theta_{h_2}$\; \nllabel{line:init_models}

\For{$t=1$ \KwTo $T$}{ \nllabel{line:loop_start}

  \textbf{Privatizer:} \nllabel{line:priv_start}\\
  
  Sample mini-batch $\{(x_j, y_j, z_{1,j}, z_{2,j}, r_j)\}_{j=1}^M$\; \nllabel{line:priv_batch}
  
  $f_j = k_{\theta_k}(x_j; r_j)$\; \nllabel{line:priv_forward}
  
  $\hat{x}_j = g_{\theta_g}(f_j, y_j)$\; \nllabel{line:recon_forward_in_priv}

  $\tilde{x}_{i,j} = h_{i,\theta_{h_i}}(f_j, z_{i,j})$ for $i=1,2$\; \nllabel{line:adv_forward_in_priv}
  \BlankLine
    
  $\mathcal{L}_{\mathrm{recon}} = \frac{1}{M} \sum_{j=1}^M d(\hat{x}_j, x_j)$\; \nllabel{line:l_recon_in_priv}
  \BlankLine
  
  $\mathcal{L}_{\mathrm{adv},i} = \frac{1}{M} \sum_{j=1}^M \ell(\tilde{x}_{i,j}, x_j)$ for $i=1,2$\; \nllabel{line:l_adv_in_priv}
  \BlankLine
    
  $\mathcal{L}_{\mathrm{adv}} = \min\{\mathcal{L}_{\mathrm{adv},1}, \mathcal{L}_{\mathrm{adv},2}\}$\; \nllabel{line:adv_min_done}

  $\mathcal{L}_{\mathrm{priv}} = -\mathcal{L}_{\mathrm{adv}} 
  + \frac{\rho}{2} \biggl( 
  \bigl( \mathcal{L}_{\mathrm{recon}} - D \bigr)^+ 
  + \bigl( D - \mathcal{L}_{\mathrm{recon}} \bigr)^+ 
  \biggr)$\; \nllabel{line:priv_loss}

  Update $\theta_k$ by gradient descent on $\mathcal{L}_{\mathrm{priv}}$\; \nllabel{line:update_k}\nllabel{line:priv_end}

  \textbf{Reconstructor:} \nllabel{line:recon_start}\\
  Sample mini-batch $\{(x_j, y_j, r_j)\}_{j=1}^M$\; \nllabel{line:recon_batch}

  $f_j = k_{\theta_k}(x_j; r_j)$, $\hat{x}_j = g_{\theta_g}(f_j, y_j)$\; \nllabel{line:recon_forward}

  $\mathcal{L}_{\mathrm{recon}} = \frac{1}{M} \sum_{j=1}^M d(\hat{x}_j, x_j)$\; \nllabel{line:l_recon_in_recon}
  \BlankLine

  Update $\theta_g$ by gradient descent on $\mathcal{L}_{\mathrm{recon}}$\; \nllabel{line:update_g}\nllabel{line:recon_end}

  \textbf{Adversaries:} \nllabel{line:adv_start}\\

  \For{$i=1$ \KwTo $2$}{ \nllabel{line:adv_loop}

    Sample mini-batch $\{(x_j, z_{i,j}, r_j)\}_{j=1}^M$\; \nllabel{line:adv_batch}

    $f_j = k_{\theta_k}(x_j; r_j)$, $\tilde{x}_{i,j} = h_{i,\theta_{h_i}}(f_j, z_{i,j})$\; \nllabel{line:adv_forward}

    $\mathcal{L}_{\mathrm{adv},i} = \frac{1}{M} \sum_{j=1}^M \ell(\tilde{x}_{i,j}, x_j)$\; \nllabel{line:l_adv_in_adv}

    Update $\theta_{h_i}$ by gradient descent on $\mathcal{L}_{\mathrm{adv},i}$\; \nllabel{line:update_h}

  } \nllabel{line:adv_loop_end}

} \nllabel{line:loop_end}

\Return $\theta_k,\theta_g,\theta_{h_1},\theta_{h_2}$\; \nllabel{line:return}

\end{algorithm}
}


Optimal solutions in Section~\ref{sec:analytical_solutions} assume specific distributions. However, real-world data may exhibit more complex, asymmetric, or unknown statistical dependencies. To accommodate broader data types, we develop a data-driven (training) procedure that jointly optimizes the privatizer, reconstructor, and adversaries. The privatizer aims to maximize the minimum adversarial loss while maintaining the reconstructor distortion below a specified threshold, and each component learns via gradient descent. 

Formally, the privatizer, reconstructor, and adversaries are the functions \(k\), \(g\), and \(h_i\), which we implement as neural networks with parameters \(\theta_k\), \(\theta_g\), and \(\theta_{h_i}\), respectively. We denote these by \(k_{\theta_k}\), \(g_{\theta_g}\), and \(h_{i,\theta_{h_i}}\). Each model operates on samples \((x_j, y_j, z_{1,j}, z_{2,j}, r_j)\) from the dataset \(\mathcal{D}\), where \(x_j\) is the private variable, \(y_j, z_{1,j}, z_{2,j}\) are side information, and \(r_j\) is an independent noise source. \(r_j\) allows for \(k_{\theta_k}\) to approximate a randomized mechanism.

We present our data-driven method in Algorithm~\ref{alg:1}. We train all components using minibatches of size \(M\) by alternating updates to the privatizer, reconstructor, and two adversaries. During the privatizer update phase (lines~\ref{line:priv_start}--\ref{line:priv_end}), its parameters \(\theta_k\) are updated while the reconstructor and adversaries operate in inference mode: their parameters \(\theta_g, \theta_{h_1}, \theta_{h_2}\) remain fixed. However, their outputs (\(\hat{x}\) from the reconstructor and \(\tilde{x}_1, \tilde{x}_2\) from the adversaries) are used to compute the privatizer loss (line~\ref{line:priv_loss}). The reconstructor's and the adversaries' networks act as surrogates for the optimal reconstructor and adversaries.

Next, during the reconstructor training phase (lines~\ref{line:recon_start}--\ref{line:recon_end}), the privatizer parameters remain fixed, but its output \(f\) is used as input to the reconstructor along with the side channel \(y\) (line~\ref{line:recon_forward}). The adversaries are not involved and remain unaffected during this stage. Similarly, during each adversary's training phase (lines~\ref{line:adv_start}--\ref{line:adv_loop_end}), the frozen privatizer provides the feature \(f\), which is combined with the corresponding side channel \(z_i\). The reconstructor is not involved in this stage, and its parameters are unchanged.

During training, the reconstructor and each adversary independently minimize their respective distortion and losses using the privatizer's sanitized output and their side information. The reconstructor minimizes \(\mathcal{L}_{\mathrm{recon}}\) (line~\ref{line:l_recon_in_recon}), while each adversary minimizes \(\mathcal{L}_{\mathrm{adv},i}\) (line~\ref{line:l_adv_in_adv}). Each component operates without constraints, aiming to approximate the optimal estimator for the current fixed privatizer.

The privatizer, in turn, minimizes the composite objective
\begin{equation}
\mathcal{L}_{\mathrm{priv}} = -\mathcal{L}_{\mathrm{adv}} 
+ \frac{\rho}{2} \left( 
\bigl( \mathcal{L}_{\mathrm{recon}} - D \bigr)^+ 
+ \bigl( D - \mathcal{L}_{\mathrm{recon}} \bigr)^+ 
\right),
\label{eq:priv_loss}
\end{equation}
where \(\mathcal{L}_{\mathrm{adv}} = \min\{\mathcal{L}_{\mathrm{adv},1}, \mathcal{L}_{\mathrm{adv},2}\}\). The second term in~\eqref{eq:priv_loss}, which penalizes deviations from the distortion threshold, is equivalent to \(\rho \cdot \bigl| \mathcal{L}_{\mathrm{recon}} - D \bigr|\). The parameter \(\rho > 0\) is the penalty weight and controls the strength of this distortion penalty~\cite{lillo1993solving}.

This formulation balances privacy and utility by maximizing adversarial loss while keeping the reconstructor distortion near the threshold \(D\). The symmetric penalty in~\eqref{eq:priv_loss} enables unconstrained training~\cite{lillo1993solving} and encourages tight satisfaction of the distortion constraint (see also \cite[Sec.~17.1]{nocedal2006numerical}). Unlike one-sided penalties, which only discourage exceeding \(D\), this design aims to get reconstructor distortion close to \(D\). Since increasing \(D\) cannot reduce adversarial loss, the privatizer is incentivized to push the distortion up to \(D\), resulting in the constraint in~\eqref{eq:optimization_problem} to hold with equality. 


\section{Simulation Results}\label{sec:simulation_results}

We evaluate our data-driven framework on both Gaussian and binary datasets. In both cases, each component (the privatizer, reconstructor, and two adversaries) is implemented as a feedforward neural network with a single hidden layer, as defined by the functions \(k_{\theta_k}\), \(g_{\theta_g}\), and \(h_{i,\theta_{h_i}}\) in Algorithm~\ref{alg:1}. The architecture consists of a fully connected \(2 \times 50\) layer with a ReLU activation function, followed by a \(50 \times 1\) output layer. The only difference is the use of a sigmoid activation after the last layer in the binary data case, versus no activation in the Gaussian data case.
To achieve a more robust convergence, we apply a set of custom techniques tailored to our training setup. Our empirical results closely match the predictions of Propositions~\ref{prop:gaussian_conditional_var} and~\ref{prop:binary_lp_solution}.


\subsection{Gaussian Data Model}

We generate a total of 12{,}000 samples \(\{(x, y, z_1, z_2)\}\) from a four-dimensional multivariate normal distribution with means \((\mu_X, \mu_Y, \mu_{Z_1}, \mu_{Z_2}) = (4, 3, 4.5, 5)\), and variances \(\sigma_X^2 = 16\), \(\sigma_Y^2 = 0.90\), \(\sigma_{Z_1}^2 = 12.25\), and \(\sigma_{Z_2}^2 = 2.25\)~\cite[Sec.~2.3]{bishop2006pattern}. The correlation coefficients are \(\rho_{XY} \approx 0.80\), \(\rho_{XZ_1} \approx 0.11\), \(\rho_{XZ_2} \approx 0.65\), \(\rho_{YZ_1} \approx 0.23\), \(\rho_{YZ_2} \approx 0.59\), and \(\rho_{Z_1Z_2} \approx 0.19\). Out of the 12{,}000 samples, 10{,}000 are used for training and 2{,}000 for testing.


Training is performed using the Adam optimizer with learning rate \(0.001\) and mini-batch size \(M = 200\). The privatizer receives a minibatch \((x, r)\), where \(r\) is standard normal noise sampled independently, and delivers a sanitized output \(f\). The reconstructor receives \((f, y)\) and produces an estimate \(\hat{x}\), while each adversary \(i = 1, 2\) receives \((f, z_i)\) and outputs \(\tilde{x}_i\). Both reconstructor distortion and adversarial loss are based on MSE.

The privatizer is trained using the loss function in~\eqref{eq:priv_loss}, with penalty weight \(\rho = 1000\), which enforces the distortion constraint~\cite{lillo1993solving} by penalizing deviations from the target distortion \(D\)~\cite[Sec.~17.1]{nocedal2006numerical}. We vary \(D\) over 30 evenly spaced values from \(D_0 = 0.005\) to \(D_{\text{max}} = \Var[X | Y] \approx 0.5502\).

For each value of \(D\), we run training trials and, in each trial, retain the final \(K = 5\) epochs whose reconstructor distortion satisfies $\mathcal{L}_{\mathrm{recon}} \in [D-\tau, D]$, where \(\tau = \frac{D_{\text{max}} - D_0}{2 \times 30}\). From these, we select the epoch with the median adversarial loss to represent the trial. A trial is accepted if, for both adversaries, the selected point (defined by reconstructor distortion and adversarial loss) has a higher adversarial loss than the corresponding finalized point from the previous distortion threshold \(D\). This enforces a non-decreasing privacy–utility tradeoff, as adversarial MSE cannot improve when more distortion is allowed. We continue collecting trials until we obtain 15 accepted pairs, each satisfying the condition that both adversarial losses do not decrease relative to the previous distortion threshold. Among these, we select the pair whose mean adversarial loss is median across all accepted pairs.

\begin{figure}[t]
    \centering    
    \input{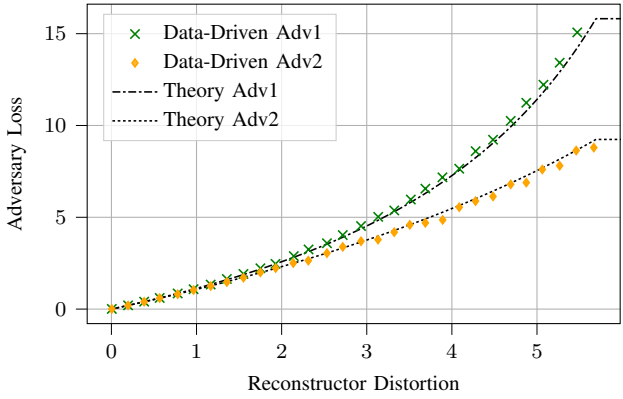}
    \caption{Gaussian data model: Data-driven versus theoretical results for adversarial loss vs.~reconstructor distortion for multiple \(D\) values.}
    \label{fig:final_distortion_plot}
\end{figure}

Figure~\ref{fig:final_distortion_plot} shows the adversarial losses as a function of the reconstructor distortion, along with the optimal solution curves (Proposition~\ref{prop:gaussian_conditional_var}). For \(D \geq D_{\text{max}} = \Var[X | Y] \approx 5.69\), the side information \(y\) solely suffices to satisfy the distortion constraint. Beyond this point, the privatizer can make $F$ independent of $X$, and the adversaries are forced to rely on their side information alone. Consequently, the solution to \eqref{eq:optimization_problem} does not change and adversarial losses then saturate at \(\Var[X | Z_1] \approx 15.81\) and \(\Var[X | Z_2] \approx 9.24\).

\begin{figure}[t]
    \centering    
    \input{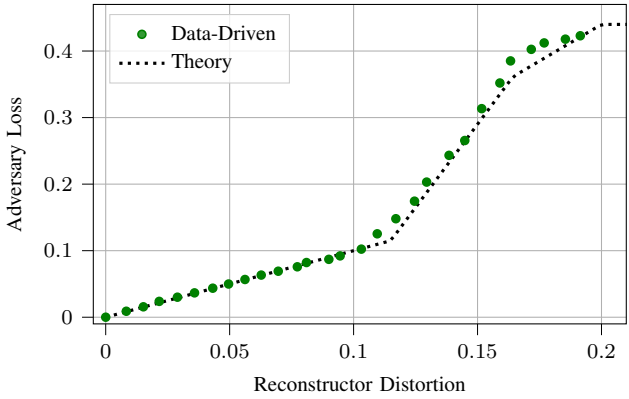}
    \caption{Binary data model: Data-driven versus theoretical result for adversarial loss vs.~reconstructor distortion for multiple \(D\) values.}
    \label{fig:binary_results}
\end{figure}


\subsection{Binary Data Model}

In the binary case, we simulate a single adversary for simplicity, as the two-adversary extension follows directly and is demonstrated in the Gaussian setting. We generate 12{,}000 samples \((x, y, z)\) with \(x \sim \text{Bernoulli}(0.54)\), using 10{,}000 for training and 2{,}000 for testing. Side information \(y\) and \(z\) are obtained by passing \(x\) through binary symmetric channels with crossover probabilities \(q_y = 0.2\) and \(q_z = 0.44\), yielding correlations \(\rho_{XY} \approx 0.6\) and \(\rho_{XZ} \approx 0.12\). The reconstructor and adversary observe \((f, y)\) and \((f, z)\), respectively, and their MAP errors based on side information alone are \(\mathcal{L}_{\mathrm{recon}} = 0.200\) and \(\mathcal{L}_{\mathrm{adv}} = 0.440\).


The privatizer receives \(x\) and outputs a probability \(\widehat{p}_x \in [0,1]\), from which a sanitized bit \(f \sim \mathrm{Bernoulli}(\widehat{p}_x)\) is sampled. The reconstructor and adversary take inputs \((f, y)\) and \((f, z)\), respectively, and produce estimates \(\widehat{x}\) and \(\tilde{x}\). Both are evaluated using empirical 0-1 loss. For the reconstructor, the distortion is given by \( \mathcal{L}_{\mathrm{recon}}(x, \hat{x}) = \frac{1}{M} \sum_{j=1}^{M} \left( x_j (1 - \hat{x}_j) + (1 - x_j)\hat{x}_j \right) \). For each adversary \(i = 1, 2\), we substitute \(\hat{x}_j\) with \(\tilde{x}_{i,j}\) to compute the corresponding loss \( \mathcal{L}_{\mathrm{adv},i}(x, \tilde{x}_i) \).

During privatizer updates, we substitute \(\widehat{p}_x\) in place of \(f\) for the reconstructor to enable gradient flow (see also \cite{bengio2013}). The adversary continues to receive \(f\), since its loss does not affect the constraint. This is a practical compromise: ideally, training would use \(f\) throughout, but doing so would block gradients from the reconstructor distortion.

All models are trained using the RAdam~\cite{liu2021varia} optimizer with a base learning rate \(\eta_0 = 0.01\), specifically, \(5\eta_0\) for the privatizer, \(2\eta_0\) for the reconstructor, and \(\eta_0\) for the adversary. The learning rate varies with distortion as \(\eta(D) = \eta_0 \left(1 + \gamma \cdot \frac{D}{D_{\max}} \right)\), with \(\gamma = -0.98\), and \(D_{\text{max}} = q_y = 0.2\). We use a batch size of \(M = 200\) and train each run for \(T = 1000\) epochs. To enforce the distortion constraint~\cite{lillo1993solving}, the privatizer is trained using the loss in~\eqref{eq:priv_loss} with penalty weight \(\rho = 1\). A symmetric penalty encourages the reconstructor distortion to stay near the threshold \(D\), while enabling unconstrained training.

We vary \(D\) over 30 evenly spaced values from \(D_0 = 0.0025\) to \(D_{\text{max}} = q_y = 0.2\). For each value of \(D\), we run training trials and, in each trial, retain the final \(K = 5\) epochs whose reconstructor distortion satisfies $\mathcal{L}_{\mathrm{recon}} \in [D-\tau, D]$, where \(\tau = \frac{D_{\text{max}} - D_0}{2 \times 30}\). From these, we select the epoch with the median adversarial loss to represent the trial. A trial is accepted if this point achieves a higher adversarial loss than the finalized point from the previous distortion threshold, ensuring a non-decreasing privacy–utility tradeoff. We continue running trials until 15 such accepted runs are collected per \(D\), and then select the one whose adversarial loss is median among the 15.

Figure~\ref{fig:binary_results} shows adversarial loss as a function of reconstructor distortion, along with the optimal solution curve (Proposition~\ref{prop:binary_lp_solution}). Similar to the Gaussian case, the solution to \eqref{eq:optimization_problem} does not change beyond \(D \geq D_{\text{max}} = q_y = 0.2\). 


\section{Conclusion}
\label{sec:conclusion}
We introduced a privacy-preserving framework in which a privatizer releases sanitized data under a constraint on a reconstructor distortion. The reconstructor and two adversaries attempt to infer the private data using their separate side information along with the sanitized data. The privatizer maximizes the minimum adversarial loss subject to the distortion constraint. 

We also proposed a data-driven minimax training procedure using neural networks. Experiments on Gaussian and binary data confirm that the learned privatizer, reconstructor, and adversaries closely match the theoretical optima, validating the approach.


\clearpage

\bibliographystyle{IEEEtran}
\bibliography{references.bib}

\begin{thebibliography}{10}
\providecommand{\url}[1]{#1}
\csname url@samestyle\endcsname
\providecommand{\newblock}{\relax}
\providecommand{\bibinfo}[2]{#2}
\providecommand{\BIBentrySTDinterwordspacing}{\spaceskip=0pt\relax}
\providecommand{\BIBentryALTinterwordstretchfactor}{4}
\providecommand{\BIBentryALTinterwordspacing}{\spaceskip=\fontdimen2\font plus
\BIBentryALTinterwordstretchfactor\fontdimen3\font minus \fontdimen4\font\relax}
\providecommand{\BIBforeignlanguage}[2]{{%
\expandafter\ifx\csname l@#1\endcsname\relax
\typeout{** WARNING: IEEEtran.bst: No hyphenation pattern has been}%
\typeout{** loaded for the language `#1'. Using the pattern for}%
\typeout{** the default language instead.}%
\else
\language=\csname l@#1\endcsname
\fi
#2}}
\providecommand{\BIBdecl}{\relax}
\BIBdecl

\bibitem{villard2010secure}
J.~Villard and P.~Piantanida, ``Secure lossy source coding with side information at the decoders,'' \emph{2010 48th Annual Allerton Conference on Communication, Control, and Computing (Allerton)}, vol.~1, pp. 733--739, 2010.

\bibitem{ekrem2011secure}
E.~Ekrem and S.~Ulukus, ``Secure lossy source coding with side information,'' \emph{2011 49th Annual Allerton Conference on Communication, Control, and Computing (Allerton)}, vol.~1, pp. 1098--1105, 2011.

\bibitem{sankar2013utility}
\BIBentryALTinterwordspacing
L.~Sankar, S.~R. Rajagopalan, and H.~V. Poor, ``Utility-privacy tradeoffs in databases: An information-theoretic approach,'' \emph{IEEE Transactions on Information Forensics and Security}, vol.~8, no.~6, p. 838–852, Jun. 2013. [Online]. Available: \url{http://dx.doi.org/10.1109/TIFS.2013.2253320}
\BIBentrySTDinterwordspacing

\bibitem{zivarifard2024secure}
H.~ZivariFard and R.~A. Chou, ``Secure source coding resilient against compromised users via an access structure,'' \emph{IEEE Journal on Selected Areas in Information Theory}, 2024.

\bibitem{shamir1979share}
A.~Shamir, ``How to share a secret,'' \emph{Communications of the ACM}, vol.~22, no.~11, pp. 612--613, 1979.

\bibitem{Blakley1899SafeguardingCK}
G.~R. Blakley, ``Safeguarding cryptographic keys,'' in \emph{1979 International Workshop on Managing Requirements Knowledge (MARK)}, 1899, pp. 313--318.

\bibitem{huang2017context}
C.~Huang, P.~Kairouz, X.~Chen, L.~Sankar, and R.~Rajagopal, ``Context-aware generative adversarial privacy,'' \emph{Entropy}, vol.~19, no.~12, p. 656, 2017.

\bibitem{kay1993statistical}
S.~M. Kay, ``Statistical signal processing: estimation theory,'' \emph{Prentice Hall}, vol.~1, pp. Chapter--3, 1993.

\bibitem{casella2024statistical}
G.~Casella and R.~Berger, \emph{Statistical inference}.\hskip 1em plus 0.5em minus 0.4em\relax CRC press, 2024.

\bibitem{cover2006elements}
T.~Cover and J.~Thomas, \emph{{Elements of information theory}}.\hskip 1em plus 0.5em minus 0.4em\relax Wiley-Interscience, 2006.

\bibitem{lillo1993solving}
W.~E. Lillo, M.~H. Loh, S.~Hui, and S.~H. Zak, ``On solving constrained optimization problems with neural networks: A penalty method approach,'' \emph{IEEE Transactions on neural networks}, vol.~4, no.~6, pp. 931--940, 1993.

\bibitem{nocedal2006numerical}
J.~Nocedal and S.~J. Wright, \emph{Numerical Optimization}, 2nd~ed.\hskip 1em plus 0.5em minus 0.4em\relax Springer, 2006.

\bibitem{bishop2006pattern}
C.~M. Bishop and N.~M. Nasrabadi, \emph{Pattern recognition and machine learning}.\hskip 1em plus 0.5em minus 0.4em\relax Springer, 2006, vol.~4, no.~4.

\bibitem{bengio2013}
\BIBentryALTinterwordspacing
Y.~Bengio, N.~Léonard, and A.~Courville, ``Estimating or propagating gradients through stochastic neurons for conditional computation,'' 2013. [Online]. Available: \url{https://arxiv.org/abs/1308.3432}
\BIBentrySTDinterwordspacing

\bibitem{liu2021varia}
\BIBentryALTinterwordspacing
L.~Liu, H.~Jiang, P.~He, W.~Chen, X.~Liu, J.~Gao, and J.~Han, ``On the variance of the adaptive learning rate and beyond,'' in \emph{8th International Conference on Learning Representations, {ICLR} 2020, Addis Ababa, Ethiopia, April 26-30, 2020}.\hskip 1em plus 0.5em minus 0.4em\relax OpenReview.net, 2020. [Online]. Available: \url{https://openreview.net/forum?id=rkgz2aEKDr}
\BIBentrySTDinterwordspacing

\end{thebibliography}

\clearpage

\appendices

\section{Proof of \cref{prop:gaussian_conditional_var}}
\label{app:gaussian}

Recall that we consider jointly Gaussian random variables \(X, F, Y, Z_1, Z_2\), with variances \(\sigma_X^2, \sigma_F^2, \sigma_Y^2, \sigma_{Z_1}^2, \sigma_{Z_2}^2\) and pairwise correlations \(\rho_{XY}, \rho_{XZ_1}, \rho_{XZ_2},\) etc. Since \((X, F)\) is jointly Gaussian, we can write \(F = aX + bR,\) where \(R \sim \mathcal{N}(0,1)\) is independent of \(X\), and \(a, b \in \mathbb{R}\) are not both zero. This yields \(\sigma_F^2 = a^2 \sigma_X^2 + b^2\) and \(\rho_{XF} = \frac{a \sigma_X}{\sqrt{a^2 \sigma_X^2 + b^2}}\). The ratio \(a^2 / b^2\) reflects the signal-to-noise ratio (SNR) in \(F\). The MMSE reconstructor distortion is \(\EE[\Var[X | F,Y]]\). For jointly Gaussian \((X,F,Y)\), the MSE of the optimal estimator is 
\[
\EE[\Var[X | F,Y]] = \frac{b^2 \sigma_X^2\bigl(1-\rho_{XY}^2\bigr)}
{b^2 + a^2\,\sigma_X^2\,\bigl(1-\rho_{XY}^2\bigr)}.
\]
This quantity decreases as \(|a/b|\) grows. When \(a/b=0\), we have that $F$ is independent of $X$ and the reconstructor distortion is \(\EE[\Var[X | Y]] = \sigma_X^2\bigl(1-\rho_{XY}^2\bigr)\). Assuming \(D \le \sigma_X^2\bigl(1-\rho_{XY}^2\bigr)\), we can require that \(\EE[\Var[X | F,Y]] = D\) and solve for $a,b$. If $D > 0$,
\[
\left(\frac{a}{b}\right)^2 = \frac{1}{D} - \frac{1}{\sigma_X^2\bigl(1-\rho_{XY}^2\bigr)}.
\]
Substituting this into the analogous conditional variances \(\EE[\Var[X | F,Z_i]]\), $i=1,2$, yields 
\[
\EE[\Var[X | F,Z_i]] = \frac{D}{\,1 \,+\,Q_i\,D\,},
\]
where
\[
Q_i =
\frac{\rho_{XZ_i}^2 - \rho_{XY}^2}{\sigma_X^2\,
\bigl(1 - \rho_{XZ_i}^2\bigr)\bigl(1 - \rho_{XY}^2\bigr)}.
\]

Finally, the corner case $D=0$ requires $b=0$ (and $a \neq 0$). In this case, both adversaries can recover $X$ with $0$ distortion too.


\section{Proof of \cref{prop:binary_lp_solution}}
\label{app:binary}

Recall that the private variable \(X \in \{0,1\}\) is drawn from a Bernoulli distribution with parameter \(p = \Pr(X = 1)\). The privatizer maps \(X\) to a sanitized variable \(F \in \{0,1\}\) using:
\[
s_0 = \Pr(F=0 | X=0),\quad
s_1 = \Pr(F=1 | X=1).
\]
Here, $s_0$ and $s_1$ are the design choices of the privatizer. The reconstructor receives side information \(Y\), generated by passing \(X\) through a binary symmetric channel (BSC) with crossover probability \(q_Y = \Pr(Y \ne X)\). Each adversary \(i \in \{1,2\}\) observes \(Z_i\), obtained independently by passing \(X\) through a BSC with crossover probability \(q_{Z_i} = \Pr(Z_i \ne X)\). The reconstructor receives the pair \((F, Y)\) and applies the MAP rule
\begin{align*}
\hat{X}(f,y) 
&= \arg\max_{x \in \{0,1\}}\, 
\Pr(F = f | X = x) \\
&\quad \cdot \Pr(Y = y | X = x)\,\Pr(X = x),
\end{align*}
while each adversary \(i \in \{1,2\}\) observes \((F, Z_i)\) and uses
\begin{align*}
\tilde{X}_i(f,z_i) 
&= \arg\max_{x \in \{0,1\}}\, 
\Pr(F = f | X = x) \\
&\quad \cdot \Pr(Z_i = z_i | X = x)\,\Pr(X = x).
\end{align*}
These MAP estimators minimize the probability of misclassification. Accordingly, the reconstructor distortion is given by
\[
\begin{aligned}
\mathcal{L}_{\mathrm{recon}}(s_0, s_1) =\; & 
\min\left\{ (1 - p)s_0(1 - q_Y),\; p(1 - s_1)q_Y \right\} + \\
& \min\left\{ (1 - p)s_0 q_Y,\; p(1 - s_1)(1 - q_Y) \right\} + \\
& \min\left\{ (1 - p)(1 - s_0)(1 - q_Y),\; p s_1 q_Y \right\} + \\
& \min\left\{ (1 - p)(1 - s_0) q_Y,\; p s_1 (1 - q_Y) \right\},
\end{aligned}
\]
and each adversarial loss \(\mathcal{L}_{\mathrm{adv}_i}(s_0, s_1)\) is defined analogously, with \(q_Y\) replaced by \(q_{Z_i}\). Then the problem \eqref{eq:optimization_problem} becomes the following piecewise-linear program:
\[
\begin{aligned}
\max_{\gamma,\, s_0,\, s_1} \quad & \gamma \\
\text{subject to} \quad 
& \mathcal{L}_{\mathrm{adv}_i}(s_0, s_1) \ge \gamma, \quad i = 1, 2, \\
& \mathcal{L}_{\mathrm{recon}}(s_0, s_1) \le D, \\
& 0 \le s_0, s_1 \le 1.
\end{aligned}
\]
The solution \((s_0^*, s_1^*)\) characterizes the privacy-utility tradeoff by maximizing the minimum adversarial loss subject to the distortion constraint.

\end{document}